\documentclass[%
reprint,
superscriptaddress,
amsmath,amssymb,
aps,
prl,
]{revtex4-2}
\pdfoutput=1
\usepackage{graphicx}% Include figure files
\usepackage{dcolumn}% Align table columns on decimal point
\usepackage{bm}% bold math
\usepackage[caption=false]{subfig}
\usepackage{microtype}
\usepackage{braket}
\usepackage{xcolor}
\usepackage[section]{placeins}

\begin{document}

\newcommand*\lierf{\texorpdfstring{LiErF${}_4$}{LiErF4}}
\newcommand*\liyf{LiYF${}_4$}
\newcommand*\ercl{\texorpdfstring{ErCl${}_3\cdot$6H${}_2$O}{ErCl3.6H2O}}
\newcommand*\eucl{EuCl${}_3\cdot$6H${}_2$O}
\newcommand\Tstrut{\rule{0pt}{2.6ex}}         % = `top' strut
\newcommand\Bstrut{\rule[-0.9ex]{0pt}{0pt}}   % = `bottom' strut

\preprint{APS/123-QED}

\title{Negative refractive index in dielectric crystals containing stoichiometric rare-earth ions}% Force line breaks with \\

\author{Matthew C. Berrington}
\affiliation{Research School of Physics, Australian National University, Canberra, ACT, Australia}
\author{Henrik M. Rønnow}
\affiliation{Laboratory for Quantum Magnetism, Institute of Physics, École Polytechnique Fédérale de Lausanne, CH-1015 Lausanne, Switzerland}
\author{Matthew J. Sellars}
\affiliation{Research School of Physics, Australian National University, Canberra, ACT, Australia}
\author{Rose L. Ahlefeldt}
\affiliation{Research School of Physics, Australian National University, Canberra, ACT, Australia}

\date{\today}% It is always \today, today,

\begin{abstract}
We investigate the prospect of achieving negative permittivity and permeability at optical frequencies in a dielectric crystal containing stoichiometric rare-earth ions. We derive the necessary transition linewidth, ion density and electric and magnetic oscillator strengths using a simplified model of non-interacting dipoles. We identify Erbium crystals in a magnetically ordered phase as the most promising material to meet these conditions, and describe initial optical measurements of two potential candidates, \ercl{} and ${}^7$\lierf{}, which display linewidths of 3~GHz and 250~MHz, respectively. The properties of ${}^7$\lierf{} satisfied our criterion for negative permeability.
\end{abstract}

\maketitle
%\tableofcontents

\section{Introduction}
Negative refractive index materials exhibit unusual optical phenomena such as negative refraction, reverse Doppler shifts and antiparallel wave and energy propagation~\cite{veselago_electrodynamics_1968}. Negative refractive indices were first identified to occur in materials with simultaneously negative real permittivity and permeability, though this has been expanded to include strongly chiral materials~\cite{Pendry_chiral} and materials with a single negative constitutive parameter~\cite{fredkin_effectively_2002}.  The most versatile path to a negative index is the original type, which this work focuses on. Since the seminal proposal for a microwave frequency negative index material in 2000~\cite{pendry_magnetism_1999,pendry_negative_2000}, negative indices have been achieved in the microwave, terahertz, and optical domains~\cite{soukoulis_past_2011}. 

%introduce double vs negative index materials
All existing negative index demonstrations at optical wavelengths use metamaterials, manufactured arrays of metallic or dielectric meta-atoms. The permittivity and permeability of the metamaterial are controlled by engineering specific magnetic and electric resonances in the meta-atoms. The meta-atom arrays are periodic with spacing less than the wavelength of the interacting light, which minimises scattering losses and allows the metamaterial to be treated as a homogeneous material with an effective permittivity and permeability~\cite{zharov_suppression_2005,gorkunov_effect_2006}. The spatial requirements are a significant engineering challenge for optical frequency negative index materials. Successful demonstrations of negative indices at visible and near-IR wavelengths use metallic and dielectric patterns of size ${\sim}100$~nm, fabricated by electron-beam lithography or focused ion beam milling~\cite{shalaev_negative_2005,garcia-meca_low-loss_2011,chettiar_dual-band_2007,dolling_negative-index_2007,valentine_three-dimensional_2008,kante_symmetry_2012}. However, this approach is restricted to surface structures several hundred nanometers thick. While metamaterial surfaces exhibit novel phenomena~\cite{jeong_emerging_2020}, effects such as perfect lensing require a 3D material~\cite{pendry_negative_2000}.

An alternative to metamaterials is to use an ensemble of atomic resonances, however, the requirements for a negative index are challenging in atomic systems~\cite{yavuz_chapter_2018}. Achieving a negative permittivity and permeability requires a strong electric and magnetic response, yet the strength of an allowed optical magnetic dipole transition is typically ${\sim}10^{-5}$ weaker than an allowed optical electric dipole transition~\cite{Foot2005}. This disparity disqualifies most optical transitions for negative refraction, as the atomic density needed to achieve a sufficient magnetic response renders the material opaque due absorption from the electric dipole transition~\cite{kastel_tunable_2007}. One exception is the $4f^N\to4f^N$ transitions in rare-earth ions, where electric dipoles transitions are parity-forbidden in the free ion, but become weakly allowed in a non-centrosymmetric crystal environment due to configuration mixing by the crystal field. This results in comparable electric and magnetic dipole strengths for certain transitions~\cite{kaminskii2013laser}. Additionally, $4f^N\to4f^N$ transitions have optical inhomogeneous linewidths as narrow as tens of megahertz at low temperatures~\cite{macfarlane_inhomogeneous_1992,macfarlane_optical_1998,thiel_rare-earth-doped_2011}, resulting in high spectral density. The high spectral density and presence of comparable electric and magnetic resonances makes rare-earth ions a promising system for exploring negative refraction in an atomic system.

Although rare-earth ions are a good candidate, the requirements of a negative index are still extreme. Theoretical studies show negative refraction is possible in doped rare-earth crystals for extremely high doping concentrations ($\gtrsim$1\%) with optical linewidths less than 1~MHz~\cite{thommen_left-handed_2006,liu_electromagnetically_2009,fu_magnetic_2013,fu_abnormal_2012}, two order of magnitude smaller than inhomogeneous broadening from dipole-dipole interactions (see Supplementary 2). Other theoretical works propose using electromagnetically induced transparency techniques in a four-level doped rare-earth crystal, but also neglects inhomogeneous broadening~\cite{zhang_quantum-interference_2012,liu_electromagnetically_2009}. A recent proposal suggests instead using a dielectric crystal containing stoichiometric terbium combined with magnetoelectric cross-coupling to a strong $4f^8 \rightarrow 4f^75d^1$ transition~\cite{buckholtz_negative_2020}. That work predicted a negative index may be possible for a feasible inhomogenous linewidth of 25~MHz, although a suitable material remains to be identified.

In this work, we propose that a more direct route to negative refraction may be possible in a dielectric crystal containing stoichiometric erbium, without the need for strong chirality. The method we propose utilises extremely high spectral density, so we consider two crystalline materials with good potential to show narrow optical transitions. The narrowest inhomogeneous lines of any rare-earth ion has been observed in ${}^7$\liyf{} lightly doped with neodymium (10~MHz~\cite{macfarlane_optical_1998}) and erbium (16~MHz~\cite{thiel_rare-earth-doped_2011}), which motivates us to study \lierf{}. The next narrowest linewidth occurs for Eu in Eu${}^{35}$Cl${}_3\cdot$6H${}_2$O (25~MHz~\cite{ahlefeldt_ultranarrow_2016}), which motivates us to study \ercl{}. We examine the possibility of a negative index in these materials using order-of-magnitude calculations of the permeability and permittivity, and present initial optical measurements of the two materials.

%The ${}^4I_{15/2} \rightarrow {}^4I_{13/2}$ transitions within the $4f^{11}$ configuration  of erbium are known to have a strong magnetic dipole and comparable electric dipole moment~\cite{li_quantifying_2014}

\section{Atomic negative refraction requirements}

Estimating the refractive index requires an effective medium theory to relate the microscopic electric (magnetic) dipole polarisability (magnetisability) to the bulk medium's permittivity (permeability). Due to interactions between individual dipoles and the coupling of electric dipoles to oscillating magnetic fields and \emph{vice versa}, developing a correct effective medium theory is non-trivial and in most cases is only performed for cubic lattices~\cite{sozio_generalized_2015,alu_first-principles_2011,kastel_local-field_2007}. We apply the model of an isotropic bicubic lattice of electric and magnetic atomic oscillators~\cite{kastel_local-field_2007}, for which the commonly used Clausius-Mossotti relation is shown to hold for the electric and magnetic responses separately. Although this model assumes a higher symmetry than our studied crystals, the model provides sufficient insight to the approximate regime required for a negative index. 

% Due to interactions between dipoles and the coupling of electric dipoles to oscillating magnetic fields and \emph{vice versa}, the relationship between microscopic electric (magnetic) dipole strengths and macroscopic permittivies (permeabilities) is non-trivial. The relationship has been solved for some lattices of electric and magnetic dipoles~\cite{sozio_generalized_2015,alu_first-principles_2011}, but not for the scheelite lattice of mixed electric and magnetic dipoles relevant to \lierf{}. 

% Since allowed atomic magnetic dipole transitions are weak, a typical dielectric has a negligibly magnetic response. Many optics formulae therefore assume the relative magnetic permeability is equal to one ($\mu_r\approx1$), describing optical dispersion, propagation and non-linear intensity effects by the interaction of electric fields with electric dipoles~\cite{Jackson:100964,LandauLifshitz}\textcolor{blue}{[]}.

The Clausius-Mossotti relation that relates the electric polarisability ($\alpha_e$) of a single atomic oscillator and the relative electric permittivity ($\epsilon_r$) is
\begin{align}
\frac{\rho \alpha_e}{3\epsilon_0} = \frac{\epsilon_r-1}{\epsilon_r+2}
\label{eqn:Clausius_Mossotti_relation}
\end{align}
% Magnetic analog:
% \begin{align}
% \frac{\rho \alpha_m\mu_0}{3} = \frac{\mu_r-1}{\mu_r+2}
% \end{align}

where $\rho$ is the atomic number density and $\epsilon_0$ is the vacuum permittivity. The electric polarisability can be modeled as a classical damped dipole, 
\begin{align}
% \alpha_e = \frac{i}{\hbar}\frac{|\bm{d}|^2}{\gamma_e/2-i \Delta_e}
\alpha_e = \frac{-e^2}{2 m_e \omega_e}\frac{f^{ED}}{\Delta_e+i \gamma_e/2}
\label{eqn:electric_polarisability}
\end{align}

%Magnetic analog:
% \begin{align}
% \alpha_m = \frac{i}{\hbar}\frac{|\bm{\mu}|^2}{\gamma_m/2-i \Delta_m}
% \end{align}

where $e$ and $m_e$ are the charge and mass of an electron respectively, and $\omega_e$, $f^{ED}$, $\gamma_e$ and $\Delta_e$ are the resonant frequency, oscillator strength, full-width at half-maximum and detuning of the electric dipole resonance, respectively. From Eqn~\eqref{eqn:Clausius_Mossotti_relation} and Eqn~\eqref{eqn:electric_polarisability}, the real part of the permittivity is minimised at a detuning of 
\begin{align}
% \Delta_e=\frac{3\hbar\gamma_e-2|\bm{d}|^2\rho/\epsilon_0}{6\hbar}
\Delta_e=\frac{3m_e\gamma_e\omega_e-e^2f^{ED} \rho/\epsilon_0}{6m_e\omega_e}
\end{align}
and will be negative when
\begin{align}
% 1<\frac{1}{\hbar\epsilon_0}\left(\frac{|\bm{d}|^2 \rho}{\gamma_e}\right)
1<\frac{e^2}{2m_e \epsilon_0 \omega_e}\left(\frac{f^{ED} \rho}{\gamma_e}\right)
\label{eqn:negative_permittivity}
\end{align}
Following the same procedure with magnetic analogs of Eqns~\eqref{eqn:Clausius_Mossotti_relation}-\eqref{eqn:negative_permittivity}, we obtain an order-of-magnitude requirement for a negative permeability, 
\begin{align}
% 1<\frac{\mu_0}{\hbar}\left(\frac{|\bm{\mu}|^2 \rho}{\gamma_m}\right)
1<\frac{e^2}{2m_e \epsilon_0 \omega_m}\left(\frac{f^{MD} \rho}{\gamma_m}\right)
\label{eqn:negative_permeability}
\end{align}
where $f^{MD}$, $\omega_m$ and $\gamma_m$ are the oscillator strength, frequency and full-width at half-maximum of the magnetic dipole resonance, respectively. The term in the brackets of Eqn~\eqref{eqn:negative_permittivity} and Eqn~\eqref{eqn:negative_permeability} are proportional to the spectral density of the electric and magnetic dipole resonances, respectively. If an atomic transition is able to simultaneously satisfy Eqn~\eqref{eqn:negative_permittivity} and Eqn~\eqref{eqn:negative_permeability}, then the medium will have a negative index for some frequencies near the transition.
% $\gamma\sqrt{\left(\frac{\mu_0}{\hbar}\frac{|\bm{\mu}|^2 \rho}{\gamma_m}\right)^2-1}$$

We now identify which rare-earth ions can satisfy Eqn~\eqref{eqn:negative_permeability} and therefore achieve a negative permeability. We make this identification without considering particular crystal hosts, which is possible as the magnetic dipole oscillator strengths of $4f^N\rightarrow 4f^N$ transitions change little between hosts~\cite{dodson_magnetic_2012}. We consider transitions from electronic ground states in the near-infrared to near-UV domain; at lower energies transition linewidths are strongly non-radiatively broadened, while at higher energies most crystalline hosts are opaque. We consider only ions with a spin zero nuclear isotope, as hyperfine structure would dilute the spectral density. The extreme requirement of Eqn~\eqref{eqn:negative_permeability} means that few transitions can satisfy it even when bestowing the transition with the minimum inhomogenous linewidth observed in \emph{any} rare-earth system ($\gamma=10$~MHz, Nd:YLiF${}_4$~\cite{macfarlane_optical_1998}) and the maximum theoretical ion density ($\rho=3.5\times10^{28}$~m${}^{-3}$, rare-earth nitride). An exhaustive list of such transitions was made using the free-ion magnetic dipole oscillator strengths of Dodson and Zia~\cite{dodson_magnetic_2012}, and is shown in Table~\ref{tab:potential_transitions}. The best candidate transition is the ${}^4I_{15/2}\to{}^4I_{13/2}$ transition of ${}^{166}$Er, ${}^{168}$Er and ${}^{170}$Er.

The ${}^4I_{15/2}\to{}^4I_{13/2}$ of erbium has two properties that further justifies its suitability for a negative refractive index material. Firstly, it has the second narrowest optical linewidth observed in any solid~\cite{thiel_rare-earth-doped_2011}. Secondly, comparable magnetic and electric dipole oscillator strengths have been observed in several erbium doped crystalline materials~\cite{weber_probabilities_1967,weber_radiative_1968,weber_optical_1973,li_quantifying_2014,gerasimov_high-resolution_2016}. This observation indicates $f^{ED}\approx f^{MD}$, which ensures that Eqn~\eqref{eqn:negative_permittivity} can be simultaneously satisfied with Eqn~\eqref{eqn:negative_permeability}.

\begin{table}
\begin{tabular}{c|c|c|c}
     Ion & Transition & $\lambda$~(nm) & $f^{MD}\times10^8$ \Tstrut\Bstrut\\
     \hline 
     Er${}^{3+}$ & ${}^4I_{15/2} \rightarrow {}^4I_{13/2}$ & 1528 & 31.14\Tstrut\\
     Yb${}^{3+}$ & ${}^2F_{7/2} \rightarrow {}^2F_{5/2}$ & 976 & 17.76\\
     Dy${}^{3+}$ & ${}^6H_{15/2} \rightarrow {}^4I_{15/2}$ & 441 & 5.48\\
     Er${}^{3+}$ & ${}^4I_{15/2} \rightarrow {}^2K_{15/2}$ & 366 & 3.66\\
     Gd${}^{3+}$ & ${}^8S_{7/2} \rightarrow {}^6P_{7/2}$ & 307 & 4.28\\
     Nd${}^{3+}$ & ${}^4I_{9/2} \rightarrow {}^2H_{9/2}$ & 822 & 1.25\\
     Sm${}^{3+}$ & ${}^6H_{5/2} \rightarrow {}^4G_{5/2}$ & 552 & 1.73\\
     Gd${}^{3+}$ & ${}^8S_{7/2} \rightarrow {}^6P_{5/2}$ & 301 & 2.42\\
     Dy${}^{3+}$ & ${}^6H_{15/2} \rightarrow {}^4H_{13/2}$ & 295 & 1.41\\
\end{tabular}
\caption{List of all free rare-earth ion transitions that have $\lambda<2500$~nm, occur in ions with a stable zero nuclear spin isotope, and have an oscillator strength that satisfies Eq~\eqref{eqn:negative_permeability} when $\gamma=10$~MHz and $\rho=3.5\times10^{28}$~m${}^{-3}$.}
\label{tab:potential_transitions}
\end{table}

% \begin{table}
% \begin{tabular}{c|c|c|c}
%      Ion & Transition & $\lambda$~(nm) & $|\bm{\mu}|$~($\mu_B$) \Tstrut\Bstrut\\
%      \hline 
%      Er${}^{3+}$ & ${}^4I_{15/2} \rightarrow {}^4I_{13/2}$ & 1528 & 0.63\Tstrut\\
%      Yb${}^{3+}$ & ${}^2F_{7/2} \rightarrow {}^2F_{5/2}$ & 976 & 0.38\\
%      Dy${}^{3+}$ & ${}^6H_{15/2} \rightarrow {}^4I_{15/2}$ & 441 & 0.14\\
%      Er${}^{3+}$ & ${}^4I_{15/2} \rightarrow {}^2K_{15/2}$ & 366 & 0.11\\
%      Gd${}^{3+}$ & ${}^8S_{7/2} \rightarrow {}^6P_{7/2}$ & 307 & 0.10\\
%      Nd${}^{3+}$ & ${}^4I_{9/2} \rightarrow {}^2H_{9/2}$ & 822 & 0.09\\
%      Sm${}^{3+}$ & ${}^6H_{5/2} \rightarrow {}^4G_{5/2}$ & 552 & 0.09\\
%      Gd${}^{3+}$ & ${}^8S_{7/2} \rightarrow {}^6P_{5/2}$ & 301 & 0.08\\
%      Dy${}^{3+}$ & ${}^6H_{15/2} \rightarrow {}^4H_{13/2}$ & 295 & 0.06\\
% \end{tabular}
% \caption{List of all free rare-earth ion transitions that have $\lambda<2500$~nm, occur in ions with a stable zero nuclear spin isotope, and have a magnetic dipole that satisfies Eq~\eqref{eqn:negative_permeability} when $\gamma=10$~MHz and $\rho=3.5\times10^{28}$~m${}^{-3}$.}
% \label{tab:potential_transitions}
% \end{table}

Having identified erbium as a strong candidate to satisfy Eqn~\eqref{eqn:negative_permeability}, we now consider what spectral density is realistically possible. The maximum theoretical erbium density in a non-metallic medium, $\rho=3.5\times10^{28}$~m${}^{-3}$, occurs in the exotic material rare-earth nitride (\emph{RE}N), but rare-earth crystals commonly used in spectroscopy such as Li\emph{RE}F${}_4$, \emph{RE}${}_2$O${}_3$ and \emph{RE}${}_2$SiO${}_5$ have rare-earth ion densities $\rho=1.4\text{ to }1.9\times10^{28}$~m${}^{-3}$~\cite{garcia_structure_1993,staritzky1956crystallographic,maksimov1970crystal}. Satisfying Eqn~\eqref{eqn:negative_permeability} with this density requires $\gamma \lesssim 1$~GHz. 

Many erbium systems have shown linewidths well below 1~GHz~\cite{rancic_coherence_2018,zhang_transparent_2017,li_optical_2020,popova_crystal_2019,thiel_rare-earth-doped_2011,phenicie_narrow_2019,gritsch_narrow_2021}; however, all these measurements have been made in doped systems with erbium concentrations of 50 parts-per-million or lower. At higher dopant concentrations, the appreciable disorder in the erbium and host lattices couples to the optical transition via ion-ion and ion-crystal-field interactions, broadening the inhomogeneous linewidth by many orders of magnitude~\cite{bottger_controlled_2008}. However, crystals with stoichiometric erbium have minimal lattice disorder and ion-ion interactions become homogeneous across all lattice sites (see Supplementary for simulations of this broadening). In fact, a 25~MHz inhomogeneous linewidth has been observed in the stoichiometric europium crystal EuCl${}_3\cdot$6H${}_2$O~\cite{ahlefeldt_ultranarrow_2016}. 
% \textcolor{blue}{[have YSO (0.005\%),Y2O3 (11.5ppm),YVO4 (30ppm),YPO4 (0.005\%),YLiF4 (0.005\%),TiO2 (4ppm),Si. Add YALO,YAG,CaWO4]}

The presence of narrow inhomogeneous linewidths in a stoichiometric europium crystal does not guarantee the equivalent in an erbium crystal. Europium has minimal magnetic broadening due to the negligible magnetic moment of its electronic singlet ground state, whereas Kramers ions like erbium have a large magnetic moment, which enables disorder in the electronic spin state to broaden the transition via magnetic dipole-dipole interactions. To avoid magnetic broadening that exceeds 1~GHz, it is necessary to remove disorder in the erbium electron spins, which can be achieved by cooling erbium crystals below their magnetic ordering temperature ($T_c$). For crystals whose sole magnetic ion is erbium, $T_c$ is typically below 1~K~\cite{bernath_massive_2022}. With the electron spin disorder removed, a spectral density able to satisfy Eqn~\eqref{eqn:negative_permittivity} and Eqn~\eqref{eqn:negative_permeability} is plausible.

Although a crystal containing stoichiometric rare-earth ions permits a high spectral density, the interactions between closely packed erbium ions may introduce additional structure due to collective excitation effects, which can lead to optical level splittings, sidebands and band structure~\cite{cone_chapter_1987}. This additional structure may inhibit negative refraction, as it represents a large deviation from the single-transition model of permittivity and permeability described above. The spectrum of collective excitations is difficult to predict without knowledge of the exact ion-ion interactions present, which motivates our empirical measurements. We investigate the optical broadening and structure of two crystalline materials containing stoichiometric erbium, to study the viability of producing optical linewidths sufficient for negative refraction without being impeded by additional optical structure. To our knowledge, no previous measurement exist of magnetically ordered stoichiometric erbium crystals with optical linewidths below 10~GHz. 

\section{Measurements of erbium crystals}

In this section we present measurements of \ercl{} and \lierf{} to understand the conditions under which the spectral density required for a negative index can be obtained, and to identify any additional structure that may inhibit negative refraction. To do this, we measure the transmission and reflection of light incident on a single face of the crystal. These two measurements are complementary: the transmission signal is zero on resonance, but provides information in the wings of the optical line, whereas the reflection signal probes the large change in electromagnetic impedance of the crystal at resonance. The reflection and transmission measurements alone cannot determine the permeability and permittivity, which requires phase-sensitive measurements~\cite{dolling_negative-index_2007}, but are sufficient to check our spectral density order-of-magnitude requirements developed in the previous section. \ercl{} did not fulfill the criteria for negative refraction under the performed experimental conditions, whereas \lierf{} did.

\ercl{} is a monoclinic crystal where erbium occupies a site with $C_2$ point-symmetry~\cite{marezio_crystal_1961}. It orders ferromagnetically below $T_c=353$~mK~\cite{beauvillain_critical_1975} and has an erbium concentration of $\rho=4.1\times10^{27}~\text{m}^{-3}$. When magnetically ordered, it has a maximum magnetic dipole oscillator strength $f^{MD}=2.2\times10^{-7}$ (see Supplementary 3), which requires $\gamma<190$~MHz to satisfy Eqn~\eqref{eqn:negative_permeability}. 
%0.42mu_B value is for 100mT along x. Isn't very sensitive

An \ercl{} crystal was grown from a saturated solution of erbium chloride in water, similar to the process described in \cite{ahlefeldt_ultranarrow_2016}, and cleaved twice along a $(100)$ plane to provide a 2.7~mm thick sample with high quality surfaces. Reflection and transmission measurements were performed with the crystal submerged in liquid helium at 2~K, above the magnetic ordering temperature, with a tunable laser incident $3^\circ$ from normal to the cleaved plane. Measurements were taken for light polarised parallel and perpendicular to the crystal $C_2$ axis (the $[010]$ direction). 

\begin{figure}[ht]
    \centering 

    \includegraphics[width=\linewidth]{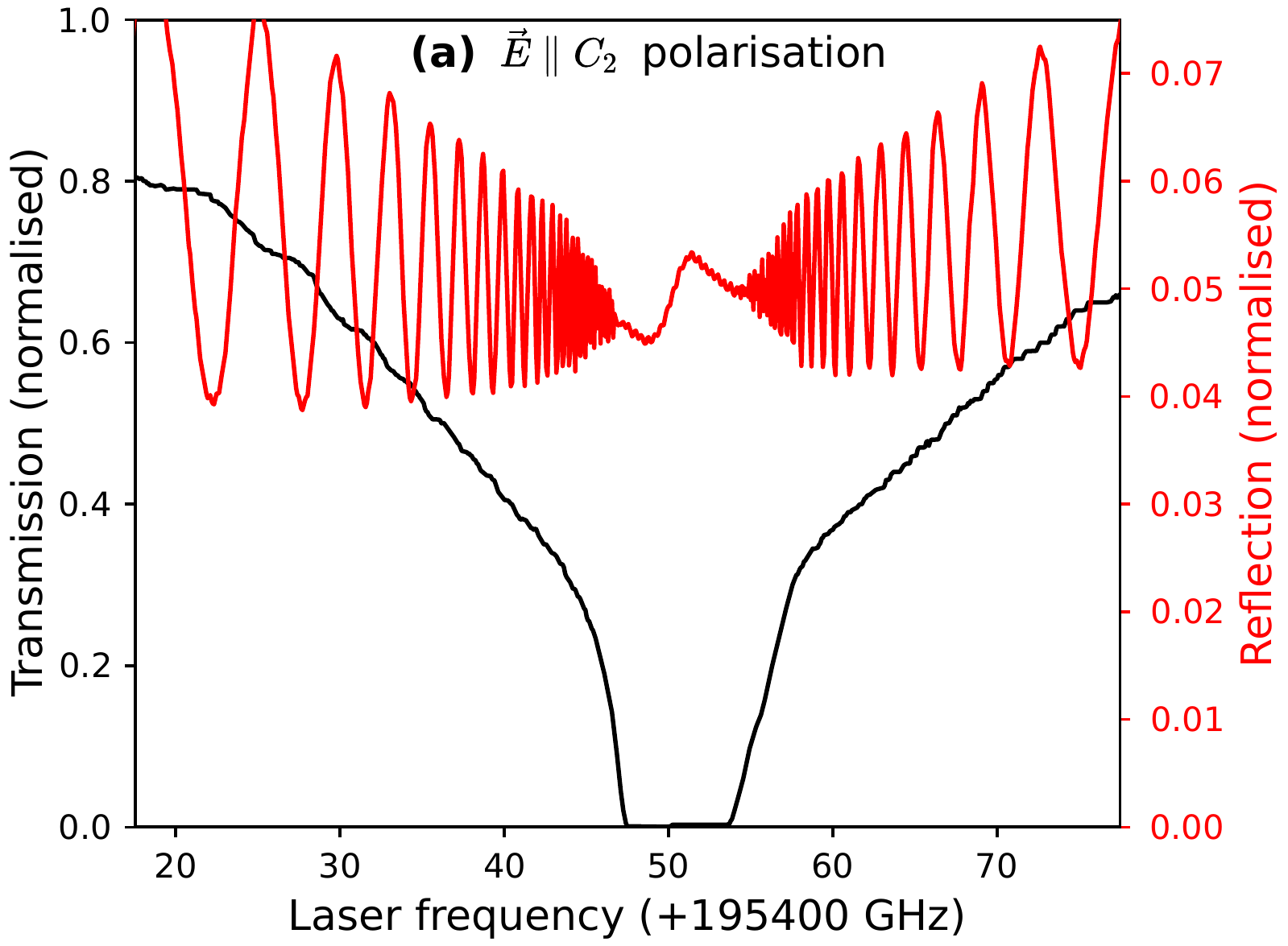}

    \includegraphics[width=\linewidth]{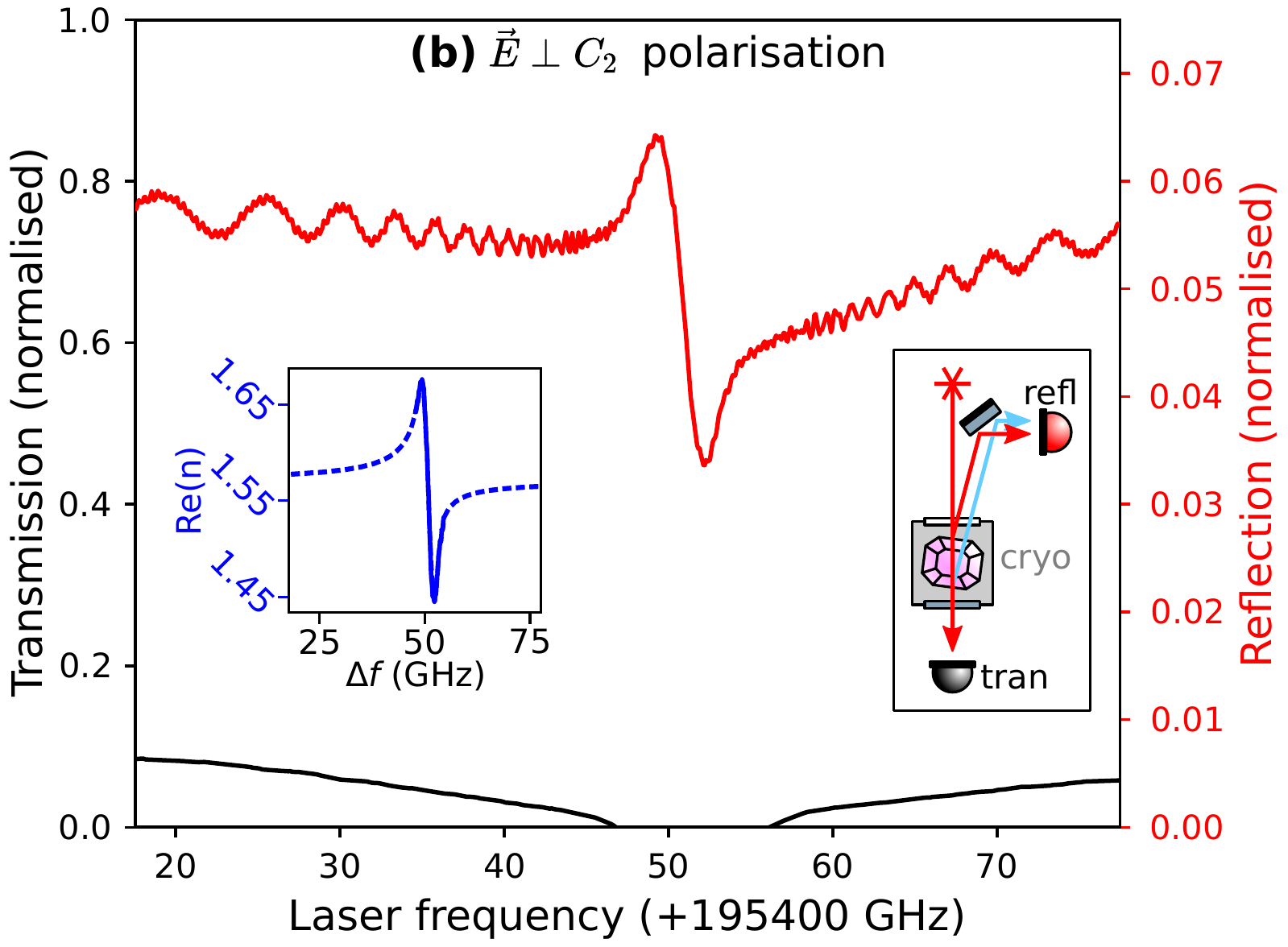}

    \caption{Transmission and reflection from a 2.7~mm thick \ercl{} crystal about the transition from the lowest ${}^4I_{15/2}$ crystal field level to the lowest ${}^4I_{13/2}$ crystal field level. Reflection has been normalised such that the mean signal in the opaque region is equal to the reflectance expected from the baseline refractive index~\cite{pabst_crystallography_1931}, and transmission has been normalised to a maximum of 1. (a) is measured with incident $\vec{E}\parallel C_2$ and $\vec{k}\perp C_2$. (b) is measured with $\vec{E}\perp C_2$ and $\vec{k}\perp C_2$. The right inset in (b) shows the measurement setup, and the left inset shows the  inferred refractive index.}
    \label{fig:reflection_measurements}
    % Reflected intensity from the face of a \ercl{} crystal at 4K\textcolor{red}{2K?}, and the absorption line recontructed using a Kramer-Kronig transformation. Talk about dashed region. Include wiggle area too?
\end{figure}

Results are shown in Figures~\ref{fig:reflection_measurements}. In the ${\sim}10$~GHz region about the centre of the resonance, no light is transmitted through the crystal and the reflected signal varies significantly. Outside the ${\sim}10$~GHz region, a beat in the reflected signal occurs due to interference of reflections from the front and back surfaces of the crystal. The beat is suppressed for $\vec{E}\perp C_2$ polarisation due to an addition background absorption.

%The broad wings of the absorption is attributed to an inhomogeneous magnetic field that broadens dislike spin transitions by tens of gigahertz, whereas the like spin transitions are broadened only a few gigahertz (see Figures~\ref{fig:reflection_measurements}a insets). This broadening is consistent with the $\mathcal{O}$(50 mT) inhomogeneous field expected from the disordered erbium magnetic moments, which 
% The frequency of the beat rapidly changes due to strong dispersion near resonance.
 
\begin{figure}[ht]
        \centering 
        \includegraphics[width=\linewidth]{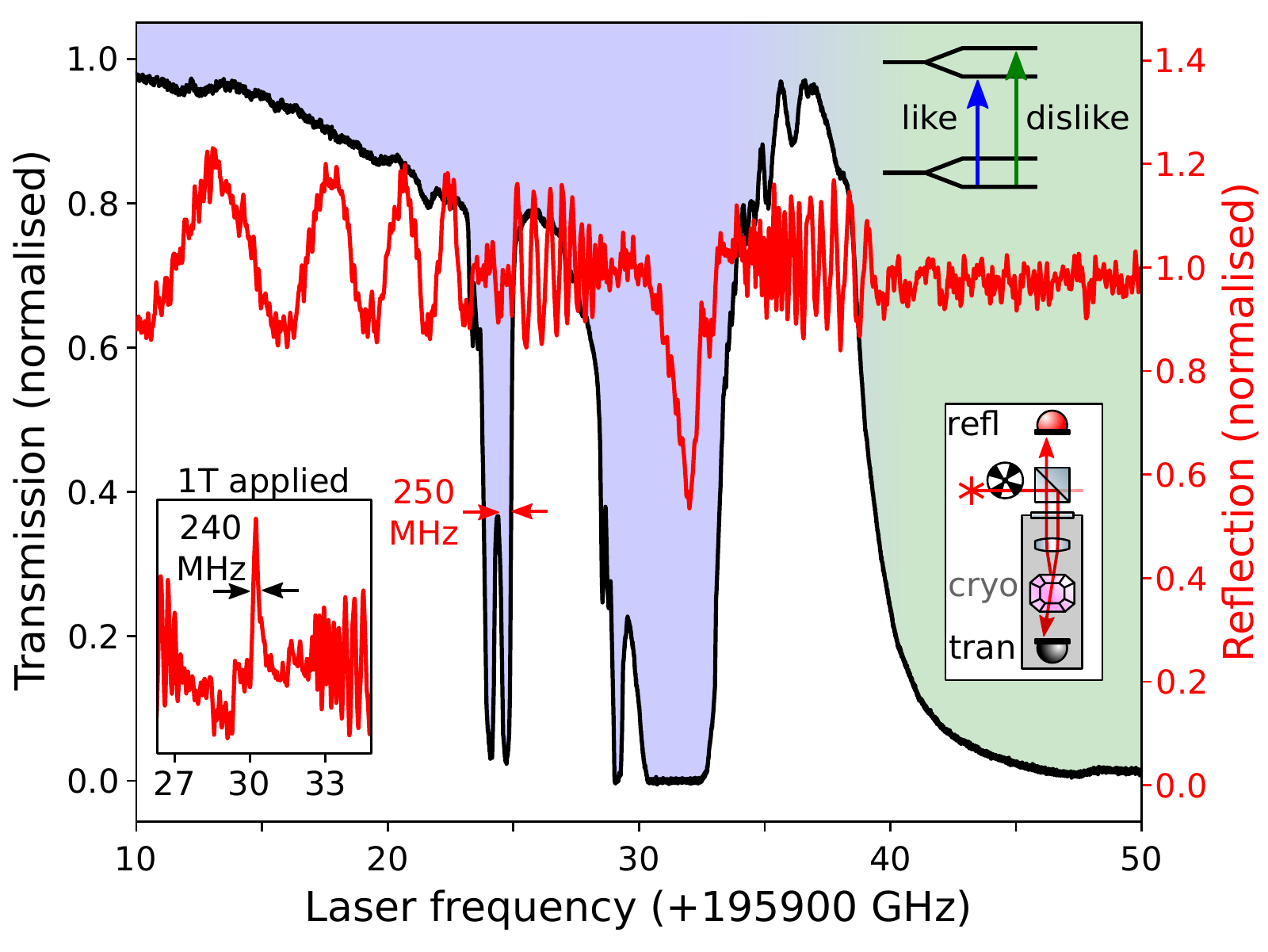}
        \caption{Transmission and reflection from a 300~{\textmu}m thick \lierf{} crystal from the lowest ${}^4I_{15/2}$ level to the lowest ${}^4I_{13/2}$ level when anchored to a 25~mK cold finger with no applied magnetic field. The right inset shows the measurement setup; a 30~nW laser of unknown polarisation was used for the reflection and transmission measurements, in combination with a chopper and lock-in amplifier. The left inset shows the reflection with a 1~T field applied. The magnetic ordering Zeeman splits the doublet, with only the ground state populated. Colour shading indicates absorption due to like$\rightarrow$like or like$\rightarrow$dislike electric spin transition.}
        \label{fig:LiErF4_reflection_measurements}
\end{figure}

The complete absorption near resonance inhibits measurement of the inhomogeneous linewidth from the transmission; however, the width of the reflected signal's dispersive curve indicates a linewidth $\gamma\approx 3$~GHz. This is inadequate for Eqn~\eqref{eqn:negative_permeability}, but was measured at 2~K$>T_c$, thus we expect large magnetic broadening to be present. We were unable to cool \ercl{} below the $353$~mK critical temperature as the crystal is efflorescent at room temperature under a vacuum, which makes it difficult to load into an ultra-low temperature refrigeration system. Nonetheless, there is a large modification to the electromagnetic impedance through the resonance, as demonstrated by the reflection intensity changing by 37\% and 9\% for the two polarisations. A Kramer-Kronig transformation of the $\vec{E}\perp C_2$ reflected signal was performed assuming, as justified in the next paragraph, the change in reflection is purely due to an electric response~\cite{silveirinha_examining_2011}. The wings of the reflected signal were reconstructed to remove the reflection from the back crystal surface~\cite{andermann_kramerskronig_1965}, giving a change in refractive index of 0.23, as shown in the Figure~\ref{fig:LiErF4_reflection_measurements}b inset. 

The reflection from the front surface follows a usual dispersion curve for $\vec{E}\perp C_2$ polarised light, yet is reversed for $\vec{E}\parallel C_2$ polarisation. We now discuss this interesting effect further. The measured reflection can be understood as interference of the reflection from a homogeneous dielectric medium with light scattered by resonant dipoles. The usual dispersion curve occurs when scattering from electric dipoles, whereas the reversed curve occurs for magnetic dipoles (see Supplementary 2). This implies that when $\vec{E}\parallel C_2$ the interaction is dominantly magnetic dipolar, whereas it is dominantly electric dipolar for $\vec{E}\perp C_2$. While in principle, the polarised magnetic dipole oscillator strengths could be calculated to verify this claim, the high sensitivity of the oscillator strengths to small magnetic fields about zero makes the calculation unprofitable, since the erbium ions will experience a range of magnetic fields due to the magnetic disorder present at 2~K (see Supplementary 3). This complication would not arise for magnetically ordered \ercl{}.

%\textcolor{red}{This observation is corroborated by group theory arguments. The true $C_2$ crystal field symmetry in \ercl{} can often be approximated as $D_{4d}$, with the $D_{4d}$ symmetry axis in the (010) plane approximately 30 degrees from [100]~\cite{couture_parametric_1984}. In this approximation, $J_z$ is a good quantum number and the lowest levels of the ${}^4I_{15/2}$ and ${}^4I_{13/2}$ crystal field multiplet both have quantum number $J_z=\frac{13}{2}$~\cite{couture_parametric_1984}, for which the magnetic dipole transition is excited by $\vec{B}\parallel D_{4d}$ and the electric by $\vec{E}\perp D_{4d}$. Given the oblique orientation of the $D_{4d}$ axis relative to the true $C_2$ axis, these selection rules imply that light polarised $\vec{E}\parallel C_2$ can excite both electric and magnetic dipole transitions, and light polarised $\vec{E}\perp C_2$ can only excite electric dipole transitions. Thus, the weaker dispersion seen for $\vec{E}\parallel C_2$ is due to the mixed character of the transition, with the magnetic dipole slightly dominating. Observation and theory therefore agree that it is possible to have dominantly electric and dominantly magnetic dipole behaviour in \ercl{} by tuning of incident polarisation. }

\ercl{} is, therefore, a promising candidate for negative refraction. It has comparable electric and magnetic dipole oscillator strength, and it has a spectral density that is only a factor of $\sim{}20$ below our negative permeability criterion, despite being magnetically disordered. 

 %It is the fully concentrated analog of Er doped in \liyf{}, in which a 16~MHz inhomogeneous linewidth has been measured\textcolor{blue}{[]}. 
\lierf{} is a tetragonal crystal where erbium occupies a site with $S_4$ point-symmetry in a scheelite configuration. It is antiferromagnetic below $T_c=375$~mK~\cite{kraemer_dipolar_2012}, has an erbium concentration of $\rho=1.4\times10^{28}~\text{m}^{-3}$ and has a magnetic dipole oscillator strength of $f^{MD} = 1.7\times10^{-7}$ (see Supplementary 3), which requires $\gamma<490$~MHz to satisfy Eqn~\eqref{eqn:negative_permeability}. A sample isotopically purified in ${}^7$Li and with natural erbium isotope abundance grown by the Czochralski method was polished close to a $(010)$ plane, 300~{\textmu}m thick.

Transmission and reflection measurements at normal incidence to the polished surface were performed in a dilution refrigerator with a mixing chamber temperature of 25~mK. A biased InGaAs photodiode on the coldfinger was used to detect transmitted light, with an optical chopper and lock-in amplifier used to increase sensitivity. The laser power was 30~nW. At higher laser powers the spectrum changed, indicating appreciable laser-induced sample heating. The low laser power resulted in a noisier reflection measurement compared to \ercl{}. Measurements were taken in both zero applied magnetic field and with a 1~T field applied along $[010]$. Polarisation of the light incident on the sample was unknown, but changes to the polarisation using waveplates outside the refrigerator had little effect on the spectra. Results are shown in Figure~\ref{fig:LiErF4_reflection_measurements}. 

The spectra of \lierf{} is more complex than \ercl, in part because the narrower linewidth reveals more structure. As the sample is magnetically ordered, an internal magnetic field at the erbium site splits the ground and excited state doublets by 31~GHz and 28~GHz respectively, with only the lower Zeeman branch of the ground state populated. We focus on the spin like$\rightarrow$like transition (see inset in Figure~\ref{fig:LiErF4_reflection_measurements}), as it is less sensitive to magnetic fields and therefore narrower.

The like$\rightarrow$like transition is fully absorbing across several gigahertz, with shoulder features on the low energy side. Most erbium isotopes are nuclear spin free, however, 23\% abundant ${}^{167}$Er has nuclear spin $I=7/2$, whose hyperfine structure is likely the origin of the absorption shoulders. The absorption lines at 195924~GHz can be interpreted as satellite features likely caused by impurities or resonant ion-ion interactions~\cite{macfarlane_chapter_1987}, and have $\gamma=250$~MHz. As the satellite features are perturbed only a few GHz from the main line, the environment of the satellite erbium must be similar to the bulk erbium. It is therefore likely that the main like$\rightarrow$like line at 195932~GHz has a similar linewidth.

The reflection near the main line is not dispersive, instead showing a pronounced dip. While a dispersive shape was observed in \ercl{}, this only occurs when the resonant contribution to the bulk refractive index is small (see Supplementary 1). Given the much narrower linewidth, here we instead expect a large change in permittivity or permeability. Despite the complex shapes possible in this regime, the width of the reflection feature bounds the resonance width. At zero field the feature width was 1~GHz, while in a 1~T external field the feature changed profile and had a $240$~MHz width (see inset in Figure~\ref{fig:LiErF4_reflection_measurements}). This matches the linewidth observed in the satellite lines.

%\textcolor{red}{Thought: at low concentration, mass disorder gives rise to structure as all nearby masses are yttrium. At high concentration, will be broadening}
% Expect F\textbf{ hyperfine structure. F spin field 5-10G (Macfarlane 1998), but different transitions have different sensitivities.
% Nd ~ 1MHz/G
% Er ~ 0.1MHz/G
% field from erbium 167 nuclear field ~ 10uT = 0.1G
% }

\section{Discussion}

In both samples, large changes in the permittivity or permeability were evident. Although the spectral density of \ercl{} did not surpass our threshold for negative permeability, this is unsurprising as the sample was magnetically disordered. Simple Monte-Carlo simulations of the resulting magnetic inhomogeneous broadening indicate ${\sim{}1}$ GHz of magnetic broadening can be expected on the like$\to$like spin transitions at zero applied field. Given the observation of a large magnetic and electric response even when disordered, cooling \ercl{} beneath its magnetic ordering temperature may achieve negative refraction.

The observed 250~MHz inhomogeneous linewidth in \lierf{} satisfies Eqn~\eqref{eqn:negative_permeability}, our criterion for negative permeability. Additionally, Gerasimov \emph{et. al.}~\cite{gerasimov_high-resolution_2016} found the electric dipole transitions contribute most of the absorption intensity over all ${}^4I_{15/2}\rightarrow {}^4I_{13/2}$ transitions in erbium-doped \liyf{}, suggesting that the negative permittivity criterion Eqn~\eqref{eqn:negative_permittivity} is also satisfied. In the current 300~{\textmu}m thick sample, strong absorption makes studying the region with a predicted negative index infeasible. To estimate the optical properties, we apply our simple model and set the electric dipole oscillator strength equal to the calculated magnetic dipole oscillator strength, $f^{ED}=f^{MD}$. In this case, the negative index region has a 420~MHz bandwidth and the minimum real refractive index occurs for a refractive index of $n=-0.97+i1.97$. This corresponds to an absorption coefficient of $\alpha=160000$~cm${}^{-1}$. At the high energy boundary to the negative index region, the absorption coefficient is $\alpha=22000$~cm${}^{-1}$, or 10\% transmission through a 1~{\textmu}m sample. To measure transmission through the negative index region we evidently require a $\mathcal{O}$({\textmu}m) thick sample, narrower optical linewidths or a loss-compensating gain mechanism.

The \lierf{} results are especially promising as the crystal was isotopically purified in lithium but not erbium, and narrower optical linewidth should be attained by purifying to a single isotope. In \liyf{} doped with 2.6~ppm erbium, broadening from the natural isotope distribution produces a ${\sim}300$~MHz linewidth~\cite{chukalina_fine_2000}, roughly matching our observation. For a fully isotopically purified crystal like ${}^7$Li${}^{168}$ErF${}_4$, the linewidth would be limited by magnetic broadening due to the disordered magnetic field from the nuclear moment of the fluorine atoms. The fluorine nuclei generate a disordered magnetic field with strength $\sim{}340$~uT~\cite{chukalina_fine_2000}, which corresponds to ${\sim}5$~MHz broadening on the like$\rightarrow$like transition. Thus, we can expect isotopically purified ${}^7$Li${}^{168}$ErF${}_4$ to achieve a significantly higher spectral density, enabling a negative refractive index with greatly reduced absorption. 

Changes in refractive indices without loss has been achieved by optically pumping multi-level atomic vapours~\cite{fleischhauer_resonantly_1992,zibrov_experimental_1996,proite_refractive_2008}. Similar techniques could be applied to erbium crystals, though technical difficulties may arise due to the low power requirements of dilution refrigerators which are needed to achieve the low temperature required for magnetic ordering. While we have focused on a two level model of erbium, the ${}^4I_{15/2}$ and ${}^4I_{13/2}$ are multiplets with other levels available for these multilevel loss compensation techniques.

% The spectral density of the magnetic could also be enhanced by increasing the transition dipole strengths. The electric and magnetic dipoles both depend on the degree of configuration mixing....The mixing is predominantly induced by the crystal field, but can also be tuned by external magnetic \textcolor{red}{and electric} fields. For example, the magnetic dipole moment for \lierf{} can be increased from $\bm{\mu}=0.27\mu_B$ to $\bm{\mu}=\textcolor{red}{\#}$ by application of \textcolor{red}{\#}~mT in the [\textcolor{red}{\#}] direction (see Supplementary). Controlling the electric dipole requires a good model of the mixing for states beyond the $4f$ orbitals, which exists for erbium doped in \liyf{}~\cite{popova_experimental_2000}, but not for the stoichiometric crystals studied. Because the spectral density scales with the dipole strength squared, tuning the dipoles with external field can provide a large enhancement.

The two erbium crystal systems studied were chosen because ultra-narrow linewidths have been measured in isostructural doped crystals. It is likely that other crystals hosts of erbium would be suitable; however, the limited studies of stoichiometric crystals have primarily focused on the energy level structure~\cite{thiel_rare-earth-doped_2011}, with little emphasis on the linewidths attainable and the required crystal growth conditions to minimise the linewidth. As such, finding other suitable crystal systems will rely on empirical measurements of seldom-grown crystals. 

Throughout this work we have assumed the simple model of Eqn~\eqref{eqn:negative_permittivity} and Eqn~\eqref{eqn:negative_permeability} is correct.  In a real atomic media this might not be the case, the main complications are the interplay between electric and magnetic dipoles belonging to the same resonance, the atomic structure that extends beyond two levels, and multipole interactions. We have also ignored the contribution of other atomic species in the lattice, which will contribute a positive real permittivity. Thus, we do not conclude  that a negative index is certain to occur, but rather that if \emph{any} atomic material is going to have a negative index without large modification of its properties, high quality stoichiometric rare-earth crystals are the best candidates.

%One of the great benefits of stoichiometric erbium crystals at a temperatures is the highly ordered ensemble attained. Stoichiometry ensures structural order, and the antiferromagnetic phase transition ensures magnetic ordering. The main disorders remaining in our sample were mass disorder from the differing erbium isotopes, and a small magnetic disorder from nuclear moments. 

% A disadvantage to generating negative refraction from a single atomic resonance with electric and magnetic characteristics is that changes in the real refractive index is always complemented by changes in the imaginary component, which manifests as absorption. Although the magnitude of the imaginary component can be much smaller than the magnitude of the real component for simultaneously strong electric and magnetic resonances~\cite{kastel_local-field_2007}, calculations based on feasible spectral densities in erbium crystals shows that imaginary refractive indices below $0.01$ are implausible. This corresponds to an absorption coefficient of $\alpha\gtrsim 800~\text{cm}^{-1}$ for the 1550~nm resonance of erbium. With this absorption, crystal thicknesses below \SI{50}{\micro\meter} are needed to get more than 1\% transmission. 

In summary, we have shown that stoichiometric erbium crystals may simultaneously have a negative permeability and permittivity at low temperature. Erbium crystals are one of the best atomic materials for achieving this condition, due to their comparable electric and magnetic dipole moments and narrow optical linewidths. To achieve a demonstrably negative index, magnetically ordered and likely isotopically pure crystals are needed. Our approach does not require any additional microfabrication or optical infrastructure, unlike metamaterials and other proposed atomic systems with a negative index.

\vspace{2cm}

\section{Acknowledgements}
% Henrik Ronnow for providing the \lierf{} sample,
We would like to thank Arne Laucht, Harish Vallabhapurapu and Chris Adambukulam for assistance operating the dilution fridge, Kieran Smith for assistance with the crystal-field model computation, and Jevon Longdell, Ilya Shadrivov and Dragomir Neshev for fruitful discussions. This work was supported by the Australian Research Council Centre of Excellence for Quantum Computation and Communication Technology (Grant No. CE170100012) and by the Australian Research Council’s Discovery Projects funding scheme (project DP210102020).

\FloatBarrier
\clearpage

\section{Supplementary 1: Resonant magnetic and electric reflections}

Here we calculate the change in reflection of an electromagnetic wave from the surface of a dielectric medium with a strong electric or a strong magnetic resonance. We will find that the change in reflected power near resonance has a different sign for electric and magnetic resonances.

\begin{figure}[h]
        \centering 
        \includegraphics[width=0.6\linewidth]{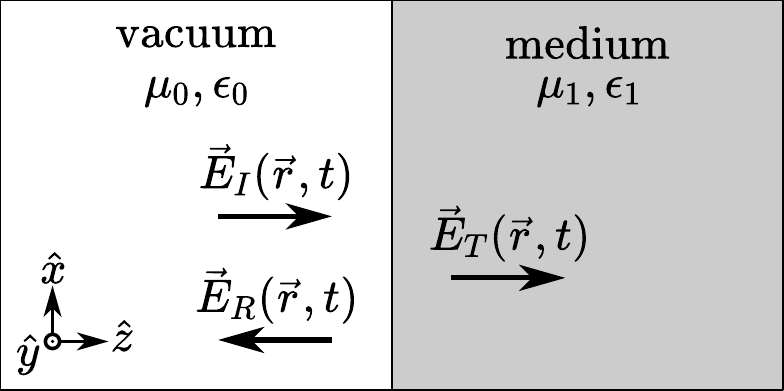}
        \caption{$\vec{E}_I(\vec{r},t)$ is a plane wave at normal incidence to the surface of a lossy homogeneous material characterised by complex permeability $\mu_1$ and permittivity $\epsilon_1$. $\vec{E}_T(\vec{r},t)$ and $\vec{E}_R(\vec{r},t)$ are the transmitted and reflected fields.}
        \label{fig:planewave}
\end{figure}

We first consider the general case of a plane wave propagating at normal incidence to the surface of a lossy material, as shown in Figure~\ref{fig:planewave}. The electric and magnetic field of the incident wave can be described by plane wave solutions to Maxwell's equations in a vacuum,
\begin{align}
\vec{E}_I(\vec{r},t) &= E_I \exp(i(\sqrt{\mu_0 \epsilon_0}\omega z-\omega t))\hat{x}\\
\vec{B}_I(\vec{r},t) &= E_I \sqrt{\mu_0 \epsilon_0} \exp(i(\sqrt{\mu_0 \epsilon_0}\omega z-\omega t))\hat{y}
\end{align}
where $\omega$ is the angular frequency of the plane wave.
Using boundary conditions for media with no free charges nor free currents, the reflected wave is
\begin{align}
\vec{E}_R(\vec{r},t) &= E_R \exp(i(-\sqrt{\mu_0 \epsilon_0}\omega z-\omega t))\hat{x}\\
\vec{B}_R(\vec{r},t) &= -E_R\sqrt{\mu_0 \epsilon_0} \exp(i(-\sqrt{\mu_0 \epsilon_0}\omega z-\omega t))\hat{y}
\end{align}
where
\begin{align}
E_R&= \frac{\sqrt{\mu_1/\epsilon_1}-\sqrt{\mu_0/\epsilon_0}}{\sqrt{\mu_1/\epsilon_1}+\sqrt{\mu_0/\epsilon_0}}E_I
\end{align}

The time-averaged power of the reflected wave is given by the Poynting vector,
\begin{align}
\langle P_R\rangle &= \frac{1}{2}\text{Re}\left(\vec{E}_R(\vec{r},t) \times \frac{1}{\mu_0}\vec{B}_R^*(\vec{r},t)\right)\\
&=-\frac{1}{2}\left|\frac{\sqrt{\mu_1/\epsilon_1}-\sqrt{\mu_0/\epsilon_0}}{\sqrt{\mu_1/\epsilon_1}+\sqrt{\mu_0/\epsilon_0}}\right|^2|E_I|^2\hat{z}
\label{eq:reflected_power}
\end{align}

We now consider specifically the medium to be a bi-cubic lattice of isotropic magnetic and electric dipoles embedded in a homogeneous dielectric with permittivity $\epsilon_{\text{bulk}}$. The magnetic response is given by magnetic analogs of Eqn~\eqref{eqn:Clausius_Mossotti_relation} and Eqn~\eqref{eqn:electric_polarisability}, resulting in
\begin{align}
\mu_1=\mu_0\frac{3(i\gamma/2+\Delta)-2\alpha_m^{\text{eff}}}{3(i\gamma/2+\Delta)+\alpha_m^{\text{eff}}}
\label{eq:mu_1}
\end{align}
where $\alpha_m^{\text{eff}}=\frac{\rho_m|\bm{\mu}|^2}{\mu_0\hbar}$ parameterises the effective magnetic polarisability. 

The electric polarisability is the sum of the electric dipole contribution and a bulk polarisability that can be treated as a constant, $P$, in a narrow frequency range near the resonance,
\begin{align}
\alpha_e = \frac{i}{\hbar}\frac{|\bm{d}|^2}{\gamma/2-i \Delta}+P
\end{align}
Using Eqn~\eqref{eqn:Clausius_Mossotti_relation},
\begin{align}
\epsilon_1 = \frac{(3+2P\rho_e/\epsilon_0)(i\gamma/2+\Delta)-2\alpha_e^{\text{eff}}}{(3-P\rho_e/\epsilon_0)(i\gamma/2+\Delta)+\alpha_e^{\text{eff}}}
\end{align}
where $\alpha_e^{\text{eff}}=\frac{\rho_e|\bm{d}|^2}{\epsilon_0\hbar}$ parameterises the effective electric polarisability. In the limit of negligible electric dipoles response, we require the material's permittivity to equal that of the bulk dielectric host, i.e.
$$\lim_{\alpha_e^{\text{eff}}\to0}\epsilon_1= \epsilon_{\text{bulk}}$$
This enforces $P=3\epsilon_0(\epsilon_{\text{bulk}}-1)/(\rho_e(\epsilon_{\text{bulk}}+2))$, which gives
\begin{align}
\epsilon_1 = \frac{9\epsilon_{\text{bulk}}(i\gamma/2+\Delta)-2\alpha_e^{\text{eff}}(2+\epsilon_{\text{bulk}})}{9(i\gamma/2+\Delta)+\alpha_e^{\text{eff}}(2+\epsilon_{\text{bulk}})}
\label{eq:epsilon_1}
\end{align}

\begin{figure}[ht]
        \centering 
        \includegraphics[width=0.9\linewidth]{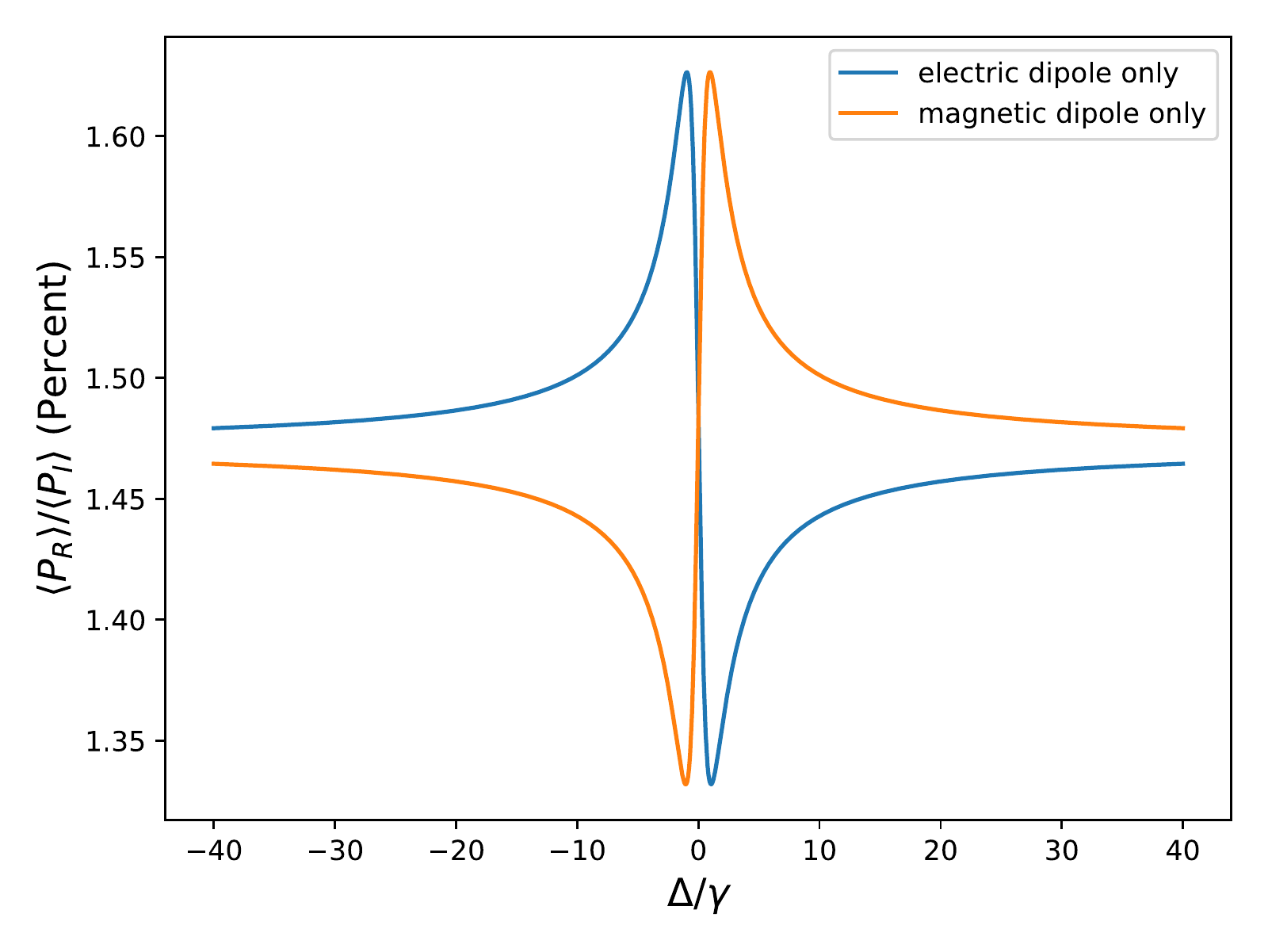}
        \caption{Reflected power when at normal incidence to an electric and a magnetic dipole material. The dielectric permittivity was set to $\epsilon_{\text{bulk}}=2$, and the values of $\alpha_e^{\text{eff}}$ and $\alpha_m^{\text{eff}}$ were such that the reflected power changed by 10\% about the resonances. The change in reflected power has opposite sign for the electric and magnetic dipole materials.}
        \label{fig:B_vs_E_reflections}
\end{figure}

The reflected signal can now be obtained using Eqn~\eqref{eq:reflected_power}, Eqn~\eqref{eq:mu_1} and Eqn~\eqref{eq:epsilon_1}. For a material with negligible magnetic dipole response, $\alpha_m^{\text{eff}}=0$, whereas when the electric dipole response is negligible, $\alpha_e^{\text{eff}}=0$. The reflected power for these two regimes is plotted in Figure~\ref{fig:B_vs_E_reflections}. Importantly, the change in reflection near resonance has the opposite sign for the dominantly magnetic and the dominantly electric dipole materials.

\begin{figure}[ht]
        \centering 
        \includegraphics[width=0.9\linewidth]{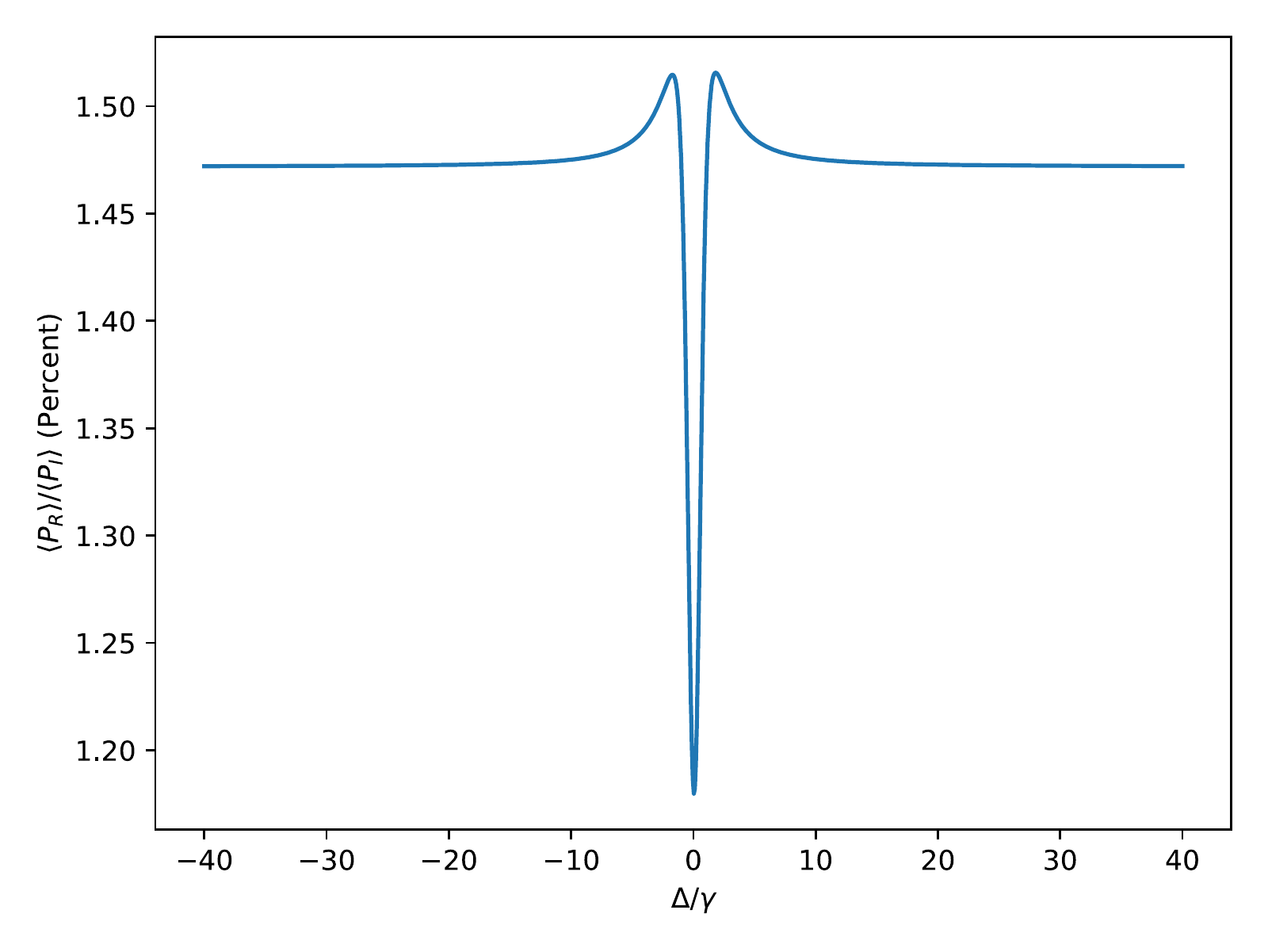}
        \caption{Reflected power when at normal incidence to a material with equal electric and magnetic dipole polarisability. The dielectric permittivity was set to $\epsilon_{\text{bulk}}=2$, and $\alpha_e^{\text{eff}}=\alpha_m^{\text{eff}}$ with a value 10 times larger than in Figure~\ref{fig:B_vs_E_reflections}}
        \label{fig:equal_magnetic_electric_reflections}
\end{figure}

The frequency dependence of the reflected signal do not follow a dispersive curve shape when the electric and magnetic dipoles polarisabilities are nearly equal, or when the change in electromagnetic impedance is similar to the off-resonant bulk contribution. An example of such a case is shown in Figure~\ref{fig:equal_magnetic_electric_reflections}, where the effective magnetic and electric polarisabilities are equal.

% \FloatBarrier
\clearpage

\section{Supplementary 2: magnetic broadening}
One source of inhomoeneous broadening to the optical linewidths in solids containing rare-earth ion is disorder in the magnetic environment. Commonly, the source of the magnetic disorder within the crystal is the rare-earth ions themselves, especially for Kramers ions like erbium which possess a large electronic dipole moment. We distinguish two types of broadening. Firstly, at finite temperatures not all dipoles will occupy the same state, which can be considered in a  classical model as disorder in the direction of the dipole. The fluctuations in the directions of dipole moments result in a different local magnetic environment for each erbium. The second type of broadening is positional disorder, where for non-stoichiometric crystals, there is disorder in the positions occupied by a particular rare-earth ion. 

In principle, the direction disorder can be supressed by applying a magnetic field when at low temperatures, such that all rare-earth ions occupy the same state. On the other hand, the positional disorder can only be controlled by changing the concentration of the different rare-earth species. Here we study the concentration dependence of the positional disorder using a Monte Carlo simulation of a LiY${}_{1-x}$Er${}_x$F${}_4$ crystal. Since erbium and yttrium have similar atomic radii, $x$ can take any value between 0 and 1 without significantly distorting the lattice. 

Out simulation lets us calculate the range of $x$ where ultra-narrow linewidths are possible. In the simulation we consider only the disorder from electronic dipole moments, neglecting the smaller nuclear dipole moments. We simulated the distributions of the magnetic fields experienced by erbium ions using the following algorithm:

\begin{enumerate}
\item A value of $x$ was chosen. For now, we consider the case when $x<0.5$ so that more yttrium is present than erbium. The other case is considered in step 6.

\item We generated a set of all possible rare-earth sites for a Li[RE]F${}_4$ lattice within a ball of radius $R$, naming the set $\{\text{RE}\}$. The radius was $R=\left(\frac{3N}{4\pi\rho x}\right)^{1/3}$, where $\rho=1.403\times10^{28}~\text{m}^{-3}$ is the density of rare-earth sites in LiErF${}_4$. This value of $R$ sets the expected number of erbium ions within the ball equal to $N=1000$.

\item We randomly chose a subset of $N-1$ sites from the set of all possible rare-earth sites $\{\text{RE}\}$. Erbium ions were assigned to this subset, also including the site at the centre of the ball, labeling the subset $\{\text{Er}\}$. Yttrium ions were assigned to all other remaining rare-earth sites, labeling the subset $\{\text{Y}\}$. Because of the judicious choice of radius, the ratio of erbium to yttrium sites is $\frac{|\{\text{Er}\}|}{|\{\text{Y}\}|}\approx\frac{x}{1-x}$. The cardinality is only approximate as the expected number of ions assumes a homogeneous density, and truncation errors will occur due to discrete lattice.
%For all calculations, $|\{\text{Y}\}|$ was within \textcolor{red}{\#} of $\frac{Nx}{1-x}$.

\item The magnetic field at the centre of the ball due to the dipole field from the other $N-1$ erbiums was calculated using the classical dipole formula
$$\mathbf{B}_0(x)=\frac{\mu_0}{4\pi}\sum_{i\in\{\text{Er}\}}
\frac{3(\mathbf{m}_i\cdot\mathbf{r}_i)\mathbf{r}_i}{|\mathbf{r}_i|^5}-\frac{\mathbf{m}_i}{|\mathbf{r}_i|^3}$$
where $\mathbf{r}_i$ and $\mathbf{m}_i$ are the positions and magnetic dipole moment of the $i$th erbium ion. The origin of the coordinate system is the erbium at the centre of the ball. To avoid pointing broadening, the magnetic dipole moments of all erbium was taken to be aligned along the crystallographic $a$ axis, such that $\mathbf{m}_i=\frac{1}{2}\mu_Bg_a\hat{a}$, where $g_a$ is the component of the ground state doublet magnetic g-tensor along the $a$ axis. 

\item Steps 3 and 4 were repeated $10^5$ times, such that a distribution of $\mathbf{B}_0(x)$ was obtained that represents the magnetic environments an erbium ion occupies.

\item Steps 1 to 4 were repeated for different values of $x$. For $0.5<x\le1$, the radius in step 2 is taken to be $$R=\left(\frac{3N}{4\pi\rho (1-x)}\right)^{1/3}$$ to ensure the expected number of yttrium within the ball equal is $N$. For $x>0.995$, we saved calculation time by using $$\mathbf{B}_0(x)=\mathbf{B}_0(1)-\mathbf{B}_0(1-x)$$ in step 4, where $\mathbf{B}_{0}(1)$ is the magnetic field at the centre of the ball when every rare-earth site is occupied by erbium. 
\end{enumerate}

\begin{figure}[ht]
        \centering 
        \includegraphics[width=0.9\linewidth]{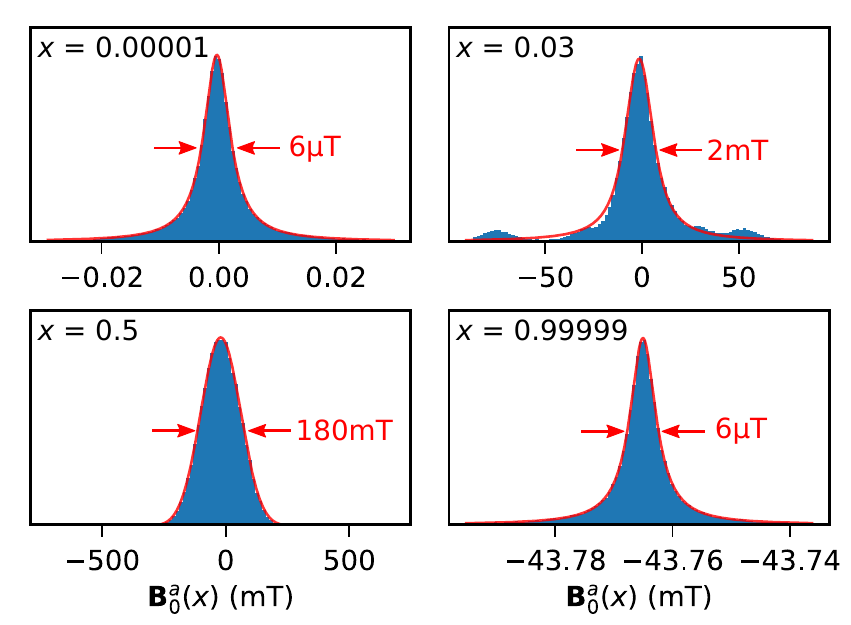}
        \caption{Distribution in the magnetic field along an $a$ axis generated from Monte-Carlo simulations. A lorentzian profile occurs for $x\lesssim0.003$ or $x\gtrsim0.997$. A gaussian profile occurs for $x\approx0.5$. In the range $0.003\gtrsim x \gtrsim 0.1$ and $0.9\gtrsim x \gtrsim 0.997$, the lineshape is complex and satellite structure is present.}
        \label{fig:magnetic_distributions}
\end{figure}

While experimenting with the algorithm parameters, it was found that $N=1000$ provided an appropriate amount disordered ions. More accurate results would occur with larger values of $N$; however requires extending the radius of the ball which rapidly increased the computation time with minimal changes in the simulation results. Simulated magnetic fields distributions are shown in Figure~\ref{fig:magnetic_distributions}. 

For extreme values of $x$ when the system is nearly stoichiometric in Er or Y, the sparse concentration of impurities causes a Lorentzian lineshape. For $x\approx0.01$ there is a significant probability that an erbium will be the nearest neighbour of the erbium at the centre of the ball, which results in a large change in the magnetic environment. The shift is large enough to create a multimodal distribution, resulting in satellite lines that are known to occur in rare-earth systems. For $x\approx 0.5$ the system is maximally disordered, resulting in a Gaussian profile. This lineshape behaviour agrees with simulations in~\cite{orth_optical_1993}.

\begin{figure}[ht]
        \centering 
        \includegraphics[width=0.9\linewidth]{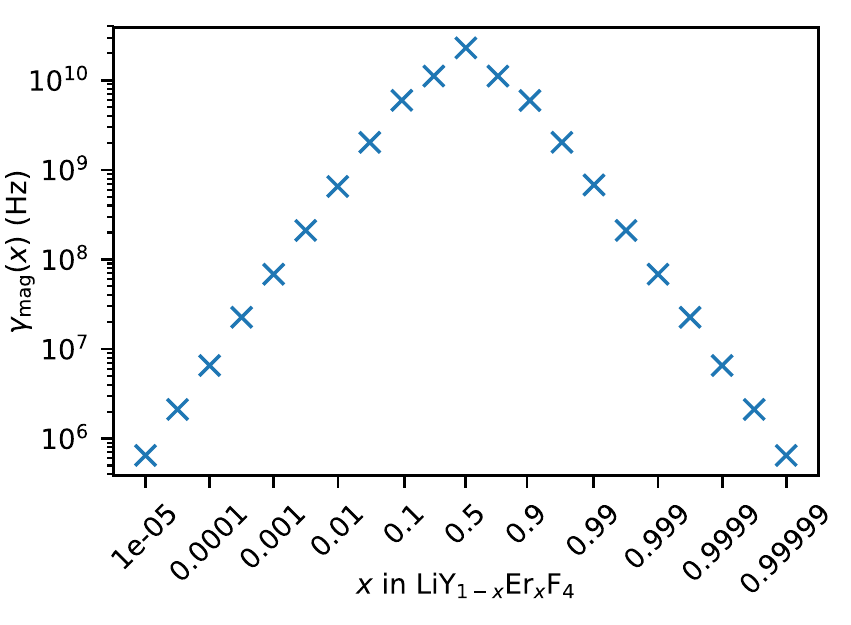}
        \caption{The calculated broadening of optical linewidths due to magnetic position broadening. For $x<0.1$ and $x>0.9$ the broadening is 660~MHz per \% impurity. The $x$-axis of this plot is logarithmic in $x/(1 - x)$.}
        \label{fig:FWHM}
\end{figure}

The lineshape of the optical transition caused by the distribution magnetic field distribution was calculated by
$$\Delta f(x) = \frac{1}{2}\mu_B (\mathbf{g}_{\text{gnd}}-\mathbf{g}_{\text{exc}})\cdot\mathbf{B_0}(x)$$
where $f\Delta(x)$ is the change optical frequency relative to a non-magnetic sample, $\mathbf{g}_{\text{gnd}}$ and $\mathbf{g}_{\text{exc}}$ are the g-tensors of the optical ground state doublets and the optical excited state doublets. The dependence of the optical linewidth, $\gamma_{\text{mag}}(x)$, on $x$ is shown in Figure~\ref{fig:FWHM}. Below $x=0.1$, the induced inhomogeneous broadening is a constant $m=650$~MHz per percent erbium. Above $x=0.9$ concentration, the broadening decreases at the same rate $m$.

\begin{figure}[ht]
        \centering 
        \includegraphics[width=0.9\linewidth]{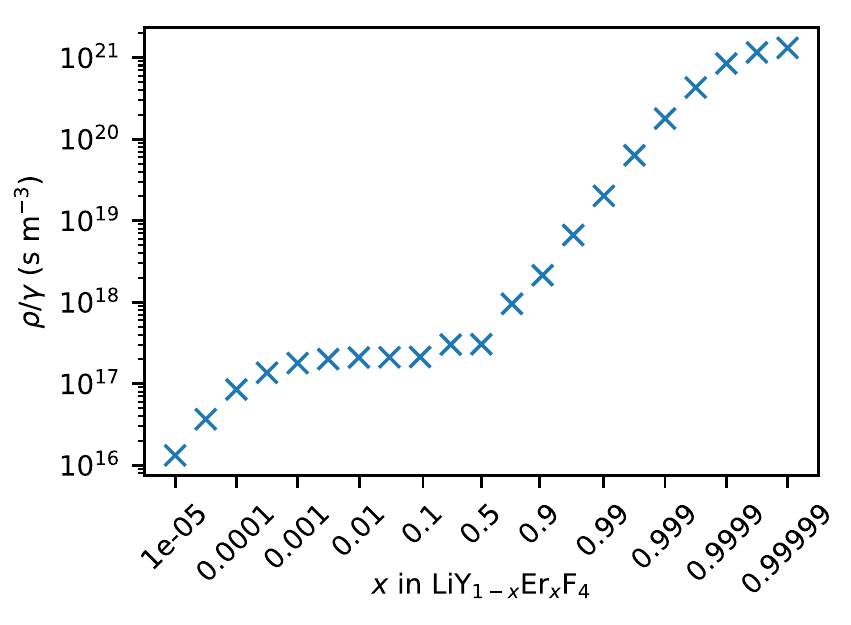}
        \caption{Erbium concentration divided by linewidth, assuming the linewidth is the sum of superhyperfine and magnetic disorder contributions. The $x$-axis of this plot is logarithmic in $x/(1 - x)$.}
        \label{fig:optimal_concentration}
\end{figure}

We now return to our motivation of finding the $x$ regime that gives the highest spectral density, which is proportional to $\rho/\gamma$. The narrowest optical inhomogeous linewidth measured in any solid is 10~MHz, in Nd:YLiF${}_4$ at a few ppm concentration~\cite{macfarlane_optical_1998}, and attributed to super-hyperfine interactions with fluorine atoms. Assuming the magnetic broadening and the superhyperfine broadening simply add,
$$\gamma = \gamma_{\text{SH}} + \gamma_{\text{mag}}(x)$$
where $\gamma_{\text{SH}}$ is the concentration independent superhyperfine broadening, the dependence of $\rho/\gamma$ is shown in Figure~\ref{fig:optimal_concentration}. The value of $\rho/\gamma$ is many orders of magnitude lower for $x<0.1$, and is bounded above by $\rho/m$ as the density and broadening both scale linearly with $x$. Only for $x>0.5$ does the scaling change, and significantly larger spectral densities become possible.

Figure~\ref{fig:optimal_concentration} demonstrates the importance of using fully-concentrated erbium crystals: below 50\% erbium concentration the spectral density is limited to $10^{-4}$ times the spectral density achievable with 100\% erbium concentration.

% NOTE to self: Using nuclear magentic dipole of Er-167 (0.565$\mu_N$), I get 360kHz FWHM at 23\% concentration. Field at fluorine lattice site when antiferromagnetic = 671mT or 615mT (two magnetic inequivalent sites). Would need to be ~1mK to freeze out fluorine spins (we don't get a frozen core effect there's no detuning from a bulk fluorine environment)

\clearpage
\section{Supplementary 3: Magnetic dipole moment calculations}

The $4f$ electrons of erbium in a host crystal can be modelled by the Hamiltonian
\begin{align}
\mathcal{H} = \mathcal{H}_{\text{FI}} + \mathcal{H}_{\text{CF}} + \mathcal{H}_{\text{Z}}
\label{eqn:total_hamiltonian}
\end{align}
where $\mathcal{H}_{\text{FI}}$ is the free-ion Hamiltonian, $\mathcal{H}_{\text{CF}}$ is the crystal field Hamiltonian and $\mathcal{H}_{\text{Z}}$ is the Zeeman interaction. Hyperfine interactions have been neglected as we are interested in zero nuclear spin isotopes of erbium. The free-ion and crystal-field Hamiltonians are defined by Eqn~(1.33) and Eqn~(1.36) respectively in~\cite{liu2006spectroscopic}. The Zeeman interaction is given by 
$$\mathcal{H}_{\text{Z}}=-\bm{\mu} \cdot \bm{H}_0$$
where $\bm{\mu}=-\mu_B(\bm{L}+2\bm{S})$ is the magnetic dipole operator.

%\textcolor{blue}{[Kieran's? Sebastian's?]
Eigenstates of Eqn~\eqref{eqn:total_hamiltonian} were found using using Mike Reid's program, permitting the calculation of the magnetic dipole matrix elements between all $4f$ levels. Magnetic dipole matrix elements were calculated for light polarised with magnetic field components of $+x-iy$, $z$, and  $-x-iy$, in accordance with Eqn~(2.14) in \cite{liu2006spectroscopic}. The $z$ direction is parallel to the crystal's symmetry axis, and $x$ and $y$ depend on the basis of the  crystal field parameters. Dipole oscillator strengths were calculated according to Eqn~(2.6) in \cite{liu2006spectroscopic}.  Results are presented below for calculations specific to \ercl{} and \lierf{}.

\subsection{\ercl{} calculations}
Free-ion Hamiltonian parameters were taken from~\cite{carnall_systematic_1989} and the crystal field parameters used were from the Er-I-Fit-R set in~\cite{karbowiak_reanalysis_2010}. These parameters have the the $x$-axis defined as the direction of maximal $g$. A refractive index of 1.57 was used~\cite{pabst_crystallography_1931}.

Two calculations were performed, the first for applied magnetic fields close to zero, relevant for the experiment shown in the results, and the second calculation with an applied field of 106.5~mT in the direction with the maximum g-tensor, which corresponds to the effective field at the erbium site when in the ferromagnetic phase~\cite{lagendijk_caloric_1973}. We present only the oscillator strengths of transitions between the lowest ${}^4I_{15/2}$ and lowest ${}^4I_{13/2}$ levels.\\

\emph{Near zero field}\\
At exactly zero magnetic field, all four transitions between the two doublets are degenerate. Upon adding the oscillator strengths in accordance with Eqn~(2.4) in \cite{liu2006spectroscopic}, the resulting polarised oscillator strengths are
\begin{align}
f^{MD}_x &= 2.4\times10^{-7}\\
f^{MD}_y &= 4.4\times10^{-7} \\
f^{MD}_z &= 3.2\times10^{-7}
\end{align}

These components are close to isotropic. However, when magnetically disordered at zero field the erbium ions will still experience a magnetic field of ${\sim{}50}$ mT due to the dipole moment of neighbouring erbium ions. This induces inhomogeneous broadening on like$\to$dislike spin transitions of ${\sim{}24}$ GHz and ${\sim{}}1$ GHz for like$\to$like spin transitions. Treating the two like$\to$like spin transitions as degenerate, the oscillator strength for an effective 1 mT field along $x$ is 
\begin{align}
f^{MD}_x &= 2.4\times10^{-7}\\
f^{MD}_y &= 4.4\times10^{-7} \\
f^{MD}_z &= 3.2\times10^{-7}
\end{align}
yet the oscillator strength for an effective 1 mT field along $z$ is 
\begin{align}
f^{MD}_x &= 2.4\times10^{-7}\\
f^{MD}_y &= 4.4\times10^{-7} \\
f^{MD}_z &= 0
\end{align}

Evidently, the $f^{MD}_z$ oscillator strength will be inhomogeneous.

\emph{Applied effective field}

The polarised magnetic dipole oscillator strengths between the lowest ${}^4I_{15/2}$ level ($Z_1$) and the lowest ${}^4I_{13/2}$ level ($Y_1$) for the applied field are

\begin{align}
f^{MD}_x &= 1.2\times10^{-7}\\
f^{MD}_y &= 2.2\times10^{-7} \\
f^{MD}_z &= 1.6\times10^{-7}
\end{align}
As no degeneracy exists in the applied field, this oscillator strength will not be inhomogenous. 

\subsection{\lierf{} calculations}
All Hamiltonian parameters were taken from~\cite{gerasimov_high-resolution_2016}, except that a magnetic field of 138~mT was applied along the crystallographic $a$ axis. This field strength corresponds to the internal magnetic field at an erbium lattice site for a spherical sample of \lierf{} in the antiferromagnetic phase. The field direction, along $a$, occurs for a antiferromagnetic domain with spins aligned $\pm b$. The crystal field parameters are defined such that $z$ corresponds to the crystallographic $c$ direction. A refractive index of 1.45 was used~\cite{barnes_temperature_1980}.

The polarised magnetic dipole oscillator strengths between the lowest ${}^4I_{15/2}$ level ($Z_1$) and the lowest ${}^4I_{13/2}$ level ($Y_1$) for the applied field are
\begin{align}
f^{MD}_x &= 3.7\times10^{-8}\\
f^{MD}_y &= 1.7\times10^{-7} \\
f^{MD}_z &= 5.2\times10^{-9}
\end{align}
The largest dipole moment occurs for light polarised with $B\parallel y$. This is equivalent to the crystallographic direction $b$.

\bibliography{main}

\end{document}